\documentclass[12pt]{revtex4-1}
\usepackage{amssymb,amsmath,fullpage}
\usepackage{graphicx}
\usepackage{color}
\usepackage{cancel}

\begin{document}

\title{The smallest matrix black hole model in the classical limit}

\author{David Berenstein$^*$, Daisuke Kawai$^\dagger$}

\affiliation {$^*$ Department of Physics, University of California at Santa Barbara, CA 93106\\
$^\dagger$ Department of Physics, Kyoto University Kyoto 606-8502, Japan.
 }

\begin{abstract} 
We study the smallest non-trivial matrix model that can be considered to be a (toy) model of a black hole. The model consists of a pair of $2\times 2$ traceless hermitian matrices with a commutator squared potential and an $SU(2)$ gauge symmetry, plus an $SO(2)$ rotation symmetry.
 We show that using the symmetries of the system, all but two of the variables can be separated. The two variables that remain display chaos and a transition from
chaos to integrability when a parameter related to an $SO(2)$ angular momentum is tuned to a critical value. We compute the Lyapunov exponents near this transition and study the critical exponent of the Lyapunov exponents near the critical point. We compare this transition to extremal rotating black holes.
 \end{abstract}

\maketitle

\section{Introduction}

Ever since the advent of the AdS/CFT correspondence \cite{Maldacena:1997re} it has become clear that many interesting quantum field theories are equivalent to theories of quantum gravity in higher dimensions. The correspondence usually entails studying the field theories at large $N$.
These can be theories in low dimensions. Particularly interesting cases occur in the BFSS matrix model \cite{Banks:1996vh}, which is a quantum mechanical theory of finitely many variables. That theory also describes some black holes in ten dimensions \cite{Itzhaki:1998dd}.

The BFSS matrix model has been subject to many quantum Monte-Carlo simulations, which have found a match between the  black hole phase and the field theory computations \cite{Anagnostopoulos:2007fw,Catterall:2008yz, Hanada:2008gy,Hanada:2008ez,Catterall:2009xn}. The most recent analysis can be found here \cite{Filev:2015hia}.
It has also been understood that chaos plays an important role in the real time thermalization properties of the ``black hole" phase. The absence of a finite temperature phase transition 
suggests that many qualitative aspects of the black hole dynamics can be understood from classical simulations of the BFSS matrix model, which has been carried out in \cite{Asplund:2011qj, Asplund:2012tg} and more recently in \cite{Gur-Ari:2015rcq}, where a spectrum of Lyapunov exponents was computed. 

 It is worthwhile to ask how much of the gravitational structure remains at very low values of $N$. Also, the computations so far have been done in the absence of angular momentum, but adding angular momentum should give rise to an interesting structure. This is because there are spinning instabilities for black holes \cite{Emparan:2003sy}. 
 
 Also, calculations of D-brane scattering \cite{Douglas:1996yp} suggest that there is a critical impact parameter (at fixed energy) which makes a scattering problem of D-branes turn into a bound state. This usually depends on an adiabatic mode at large distance separation between the branes becoming non-adiabatic. These effects create strings stretching between the branes and the branes become bound to each other. This is an interesting problem in it's own right. One can argue that the non-adiabatic behavior can be obtained with small classical perturbations for off-diagonal modes. These produce strings stretching between the D-branes that force a rescattering event. Eventually they become large to the point where they cause large back-reaction and scramble the dynamics completely (this is similar to the studies in \cite{Berenstein:2010bi, Iizuka:2013yla, Iizuka:2013kha} for a similar collision problem between D-branes). This is interpreted as the formation of a black hole.

Another advantage of working at small values of $N$ is that it becomes easier to scan over the possibilities. One might also be able to compare these kinds of situations to a direct computation of the wave functions (if the number of dimensions of the quantum mechanical problem is small enough). A simple example for the two matrix model is studied in \cite{Hubener:2014pfa} (see also \cite{Fatollahi:2015fna}), wherein a list of energies of states is obtained.

We will consider this example from the point of view of classical physics. The example arises from studying the dimensional reduction of $YM_{2+1}$ to $0+1$ dimensions. We will work
also with the $SU(2)$ model. The system only has $2$ dynamical matrices, each of them counting three real dynamical variables (and their canonical conjugates). Because of the gauge symmetry, three variables are gauged, leaving us with a dynamical system with only three dynamical variables. There is an extra $SO(2)$ symmetry that reduces the effective problem to only two dynamical variables plus their conjugates, the minimal dimension for the system to be non-integrable. The parameter that controls this reduction is the angular momentum 
of the $SO(2)$ symmetry. In this vein,  subsectors of the BMN matrix model of small dimension have been found to be chaotic \cite{Asano:2015eha}. In that case, it is the strength of the mass deformation parameter that produces islands of stability. 

At fixed energy and large angular momentum we expect  the system to be characterized by D-branes that are well separated from each other and that are orbiting each other with a number of strings stretching between  them. The number of such strings is a variable, but the occupation number counting the strings is expected to be an adiabatic invariant in these situations. On the other hand, at low angular momentum we expect large parametric resonance and  non-integrable dynamics characterized by chaos. What is interesting for us is to understand how this varies as we change the angular momentum.

The paper is organized as follows.
Sect.I\hspace{-.1em}I is devoted to the derivation of the Hamiltonian we used in the numerical simulation and in sect.I\hspace{-.1em}I\hspace{-.1em}I we discuss the properties of the chaotic behavior with Poincar\'e sections. The result on the Lyapunov exponents are presented in sect.I\hspace{-.1em}V and we conclude in sect.V.


\section{The Lagrangian and separation of variables}

Consider the 2-matrix model, where we have $X_1,X_2$ hermitian traceless $2 \times 2$ matrices. They are each in a triplet of $SU(2)$.
We want to consider the dynamical system with Lagrangian
$$
{\cal L} = \frac 12 \hbox{Tr} ( D_tX_1^2+D_t X_2^2+\frac 12[X_1,X_2]^2)
$$
This arises from the reduction of $SU(2)$ Yang Mills in $2+1$ dimensions to $0+1$ dimensions.

We can expand the $X_1,X_2$ in terms of Pauli matrices as follows
\begin{equation}
X_i = \vec x_i \cdot \vec \sigma /\sqrt{2} = x_{ji} \sigma_j/\sqrt{2}
\end{equation} 
The normalization is chosen so that in the expression $\frac12\hbox{Tr} (\dot X_i^2)=\frac 12\sum_j \dot x^2_{ji}$ has canonical kinetic terms. Any other choice of normalization can be scaled out by a rescaling of the time variable.

The collection of the two $X_{1,2}$ can be thought of as a real $3\times 2$ matrix. The gauge action is by multiplication on the left by $SU(2)\simeq SO(3)$ group elements.
The lagrangian is also symmetric under $SO(2)$ rotations of $X_1$ into $X_2$, these can be realized by multiplication on the right by an $SO(2)$ rotation. We write this as follows
\begin{equation}
\begin{pmatrix}
x_{11} & x_{12}\\
x_{21}& x_{22}\\
x_{31}&x_{32} 
\end{pmatrix} \to R_{SO(3)}\cdot  \begin{pmatrix}
x_{11} & x_{12}\\
x_{21}& x_{22}\\
x_{31}&x_{32} 
\end{pmatrix} \cdot \begin{pmatrix}\cos(\theta) & \sin(\theta)\\
-\sin(\theta)& \cos(\theta)
\end{pmatrix}
\end{equation}
Like in most holographic matrix models, the $SO(3)$ rotations are gauged. They can be written in terms of Euler angles if we want to.

What is important for us now, is that we can choose a gauge where $x_{21}=x_{31}=0$. This uses a rotation in $SO(3)$, but the $SO(2)$ rotation of the $23$ components does not affect the configuration. It is the little group associated to this gauge choice.
Similarly, we can use this freedom to choose $x_{32}=0$. This effectively reduces the number of dynamical variables from $6$ to $3$. We can still act with $SO(2)$ transformations on the right, and since they are a symmetry, we expect one conserved quantity associated to these rotations.  These preserve the $x_{31}= x_{32}$ gauge condition, but not the $x_{21}=0$ gauge. If we can separate the variables carefully in the Hamiltonian formalism, this procedure should reduce the number of degrees of freedom from $3$ to $2$, with an additional external parameter that measures the $SO(2)$ angular momentum that mixes the two matrices.  Any further reduction and the system would become integrable. 

Before we do that however, let us establish some facts in the $A_0=0$ gauge. For the system described above, the Hamiltonian can be written as 
\begin{equation}
\frac 12 \sum_{j} \vec p_{j}\!^2+ \frac 12 (\vec x_1\times \vec x_2)^2
\end{equation}
where the $\times$ symbol indicates the cross product of three vectors.
The generators of angular momentum $SU(2)$ rotations are given by
\begin{equation}
\vec L=\vec x_1\times \vec p_1+\vec x_2\times \vec p_2=\vec L_1+\vec L_2=0
\end{equation}
and correspond to the three constraints of the system that we need to specify in the initial conditions. 
All the vectors $\vec x_1, \vec x_2, \vec p_1, \vec p_2$ are orthogonal to $\vec L_1= -\vec L_2$. This is only true if the constraints are satisfied.

We will now show that the time derivative of $\vec L_1$ lies in the direction of $\vec L_1$ and therefore the motion of $\vec x_1$  is in the orthogonal plane determined by this direction.
The same argument works for $\vec x_2$. The full motion will  lie in the plane determined by $\vec x_1$ and $\vec x_2$.
To compute this time derivative, we notice that 
\begin{equation}
\partial_t(\vec L_1) = \partial_t \vec x_1 \times \vec p_1 +\vec x_1 \times \partial_t \vec p_1\label{eq:const}
\end{equation}
The first term vanishes identically by the equations of motion $\partial_t \vec x_1=\vec p_1$. 
The second term is obviously orthogonal to $\vec x_1$. The equation of motion of $\vec p_1$ is proportional to  $\vec x_2 \times( \vec x_1\times \vec x_2)$. This is also orthogonal to the $\vec L$ plane, which can be determined by any two vectors in the plane.  In this case, $\vec x_1, \vec x_2$. What we see is that the two terms in the tensor product are orthogonal to the direction of $\vec L_1$. Hence their cross product is aligned with $\vec L_1$.

Without loss of generality, we can reduce the problem to motions where $\vec L_1$ is determined by  the $12$ plane. The gauge $x_{31}= x_{32}=0$ is preserved by the equations of motion. This simplifies the analysis because we can avoid using the full Euler angles in the $SO(3)$ rotation and we can restrict ourselves to a $2\times 2 $ problem.

We will now analyze the dynamics starting from this simplification.
It is convenient to write the two vectors in the $X,Y$ plane as a $2\times 2$ matrix 
\begin{equation}
U=\begin{pmatrix} x^1_1& x^1_2\\
x^2_1& x^2_2
\end{pmatrix}
\end{equation}
The $SO(2)$ gauge transformation acts on column vectors by left multiplication, while the $SO(2)$ global symmetry acts by  right multiplication.
It is easy to see that $\det(U)\simeq \vec x_1\times \vec x_2$ and is left invariant by both such multiplications. Also, 
the following is invariant under both $SO(2)$ transformations, $\hbox{Tr}( U^T U)= \sum (x^i_j)^2= r^2$

 We will choose to write the general such matrix as follows
\begin{equation}
U= \frac 1{\sqrt 2}\begin{pmatrix} \cos(\chi) & -\sin(\chi)\\
\sin(\chi)& \cos(\chi)
\end{pmatrix}  \begin{pmatrix} 
r& r\cos \theta\\
0& r\sin\theta
\end{pmatrix}\begin{pmatrix} \cos(\phi) & -\sin(\phi)\\
\sin(\phi)& \cos(\phi)\end{pmatrix}
\end{equation}
This defines our coordinate system.

It is easy to show that it is always possible to set up the two vectors to have the same length $r^2/2$ by an $SO(2)$ rotation acting on the right. The reason for this is  that a rotation by $\pi/2$ in $\phi$ exchanges the two vectors (with a sign flip on one of them). Since the process is continuous, the difference of the length of the two vectors will go from positive to negative, so there must be a place where they are the same. 

We parametrize the misalignment by the angle $\theta$. We could have equally chosen the two vectors to be orthogonal, which would occur at the maximum or minimum value of $\vec x_1^2$,  as done in \cite{Kares:2004uk}. This produces similar results to our formulation.

Now, we write the Lagrangian in terms of the $(r, \theta, \phi, \chi)$ coordinate system. Since we choose to gauge the $\chi$ transformation, when we write the Hamiltonian we will have $p_\chi=0$, similarly, when we write the Hamiltonian we have that $p_\phi$ is conserved, so we can set it to a constant.

After computing the Jacobian for the change of variables,  the metric in the new coordinates $r, \theta,\phi,\chi$ is given by
\begin{equation}
g_{\mu\nu}\to \begin{pmatrix}
1&0&0&0\\
0&r^2/2 &  -r^2\cos(\theta)/2 &-r^2/2\\
0&  -r^2\cos(\theta)/2& r^2&r^2 \cos(\theta)\\
0&-r^2/2& r^2 \cos(\theta)&r^2
\end{pmatrix}
\end{equation}
and its inverse is
\begin{equation}
\begin{pmatrix}
1&0&0&0\\
0& 4 r^{-2} &0& 2 r^{-2}\\
0&0& r^{-2} \csc(\theta)^2& -r^{-2}\cot(\theta)\csc(\theta)\\
0&2r^{-2}& -r^{-2}\cot(\theta)\csc(\theta)&r^{-2} (2-\cos(\theta)^2)\csc(\theta)^2
\end{pmatrix}
\end{equation}

When we apply the constraint $p_\chi=0$, the kinetic term  reduces to 
\begin{equation}
\frac 12 p_r^2 +\frac 4 {r^2} p_\theta^2+ \frac {p_\phi^2}{2 r^2 \cos(\theta)^2}
\end{equation}
which is rather simple.

That is, the effective inverse metric is 
$$
g_{\mu\nu}^{-1} = \begin{pmatrix} 1& 0&0\\0&4 r^{-2}&0\\0&0& r^{-2} \cos^{-2}(\theta) \end{pmatrix}
$$

The full Hamiltonian in these coordinates is
\begin{equation}
H = \frac 12 p_r^2 +\frac 2 {r^2} p_\theta^2+ \frac {p_\phi^2}{2 r^2 \cos(\theta)^2}+\frac 14 r^4 \sin^2\theta\label{eq:Hamred}
\end{equation}
the same as the one found in \cite{Kares:2004uk}.

Notice that once $p_\theta\neq 0$, the potential becomes singular at $\cos\theta=0$, so the motion in $\theta$ is constrained to the $-\pi/2, \pi/2$ range. Similarly, the motion never reaches $r=0$.

Because the equation of motion of $p_\phi$ is given by $\dot p_\phi=0$, we find that we can treat it as a constant and we only need to evolve the two variables $r, \theta$ and their conjugate variables, $p_r, p_\theta$.

\begin{figure}[ht]
\begin{center}
\includegraphics[width=5cm]{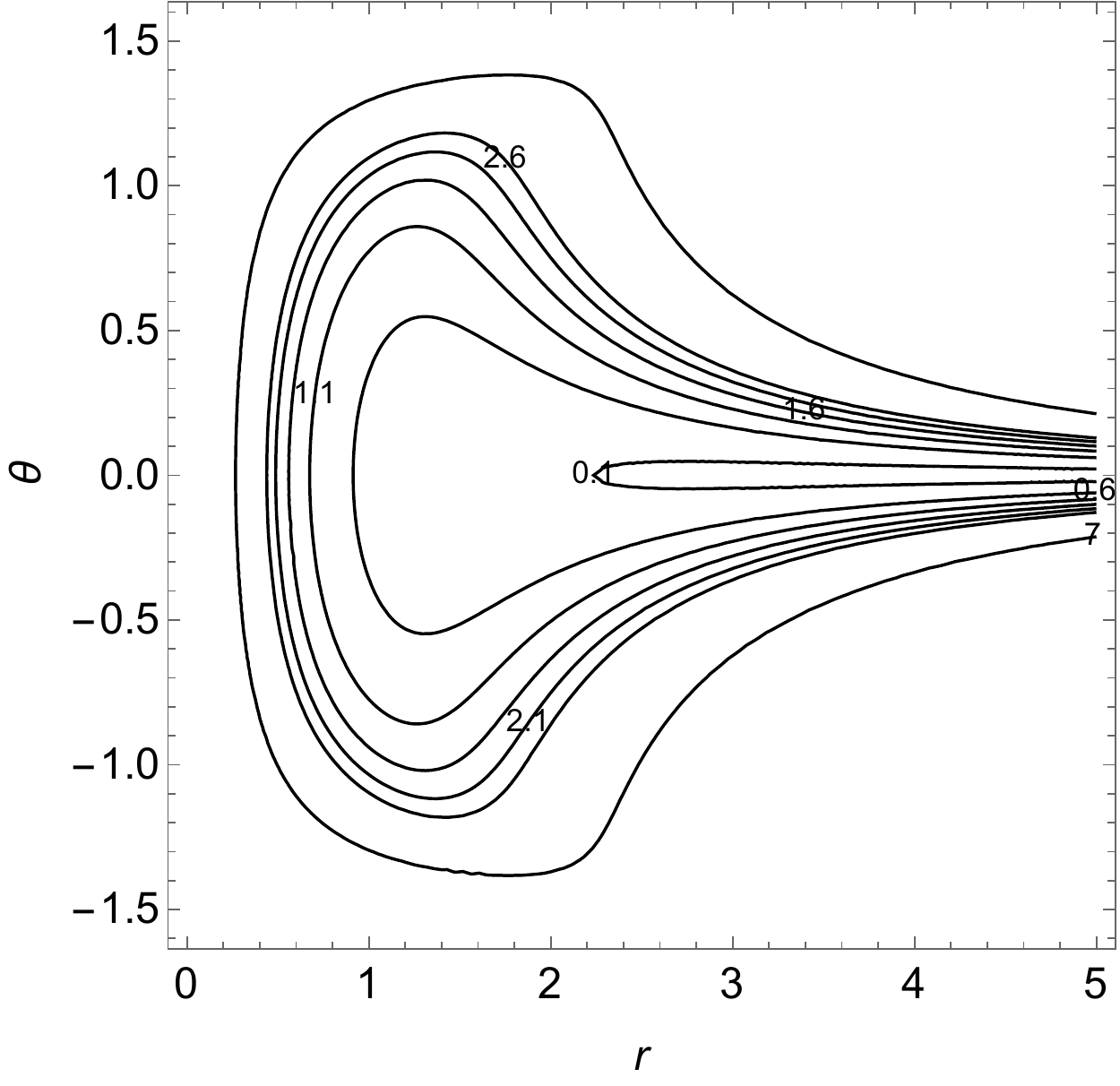}
\caption{Equipotential surfaces at $p_\phi=1$ at values $V_{pot}=0.1, 0.6,  1.1, 1.6, 2.1, 2.6, 7$} \label{fig:pot}
\end{center}
\end{figure}

It is instructive at this stage to draw a map of the potential. This is shown in figure \ref{fig:pot}.
What is important for us is the general structure of the potential. For low energy and fixed angular momentum there is always a needle shape region that extends to 
infinity at $\theta\simeq 0$. This is the region where the commutator of the two matrices vanishes. This is a flat direction in these kinds of matrix models. 

From the point of view of numerics, trajectories that take a long excursion in the needle region take a  long time to compute, but the motion in $\theta$ is expected to be adiabatic in this region. Therefore nothing much happens during these excursions. Quantum mechanically, we know that these directions will be lifted. At large $r$, we can treat the variable $\theta$ as a harmonic oscillator. The term with $r^4 \sin^2(\theta) \simeq r^4 \theta^2$ dominates the potential. The kinetic energy will be $ 2 p_\theta^2/r^2 $. The effective frequency of the $\theta$ direction is $\omega^2_{eff} \simeq r^4/r^2 =r^2$, so the effective correction to the Hamiltonian will be 
$\delta H\propto \hbar r$. We choose to modify the potential this way with a small $\hbar$. This correction mostly affects the needle region, where $r$ can become large. In the numerics 
we set $\hbar \leq 0.1$ at $E=1.0$.

The improved $\hbar$-corrected Hamiltonian in these coordinates is
\begin{equation}
H = \frac 12 p_r^2 +\frac 2 {r^2} p_\theta^2+ \frac {p_\phi^2}{2 r^2 \cos(\theta)^2}+\frac 14 r^4 \sin^2\theta+\hbar r\label{eq:Hamredcorr}
\end{equation}
where $\hbar$ is a parameter.

Also, with this correction, the potential now has a minimum at $\theta=0$ and $r\propto p_\phi^{2/3} \hbar ^{-1/3}$. This gives a bound on the energy $E > O(1) \hbar^{2/3} p_\phi^{2/3}$
that scales with a power of the angular momentum (for a spectrum of the quantum model, see \cite{Hubener:2014pfa}). 

\section{Chaos}

It is well know that the Hamiltonian \eqref{eq:Hamred} is fully chaotic for $p_\phi=0$ (this means that there are no KAM tori\cite{Ko,Ar,Mo}). This model and related models have been studied in 
\cite{Sav,Chirikov:1981cm,Aref'eva:1997es,Matinyan:1981ys} where it was shown that they are chaotic.
 This is the same dynamics as the Hamiltonian given by
\begin{equation}
\frac{p_x^2}{2}+\frac{p_y^2}{2}+\lambda x^2 y^2
\end{equation}
with $x= r\sin(\theta/2)$, $y=r\cos(\theta/2)$. What we want to understand is the presence or absence of chaos as we modify the angular momentum $p_\phi$ at fixed energy.
As we modify $p_\phi$ we see that the system evolves from being chaotic to a system that is not.

\begin{figure}[ht]
\begin{center}
\includegraphics[width=7cm]{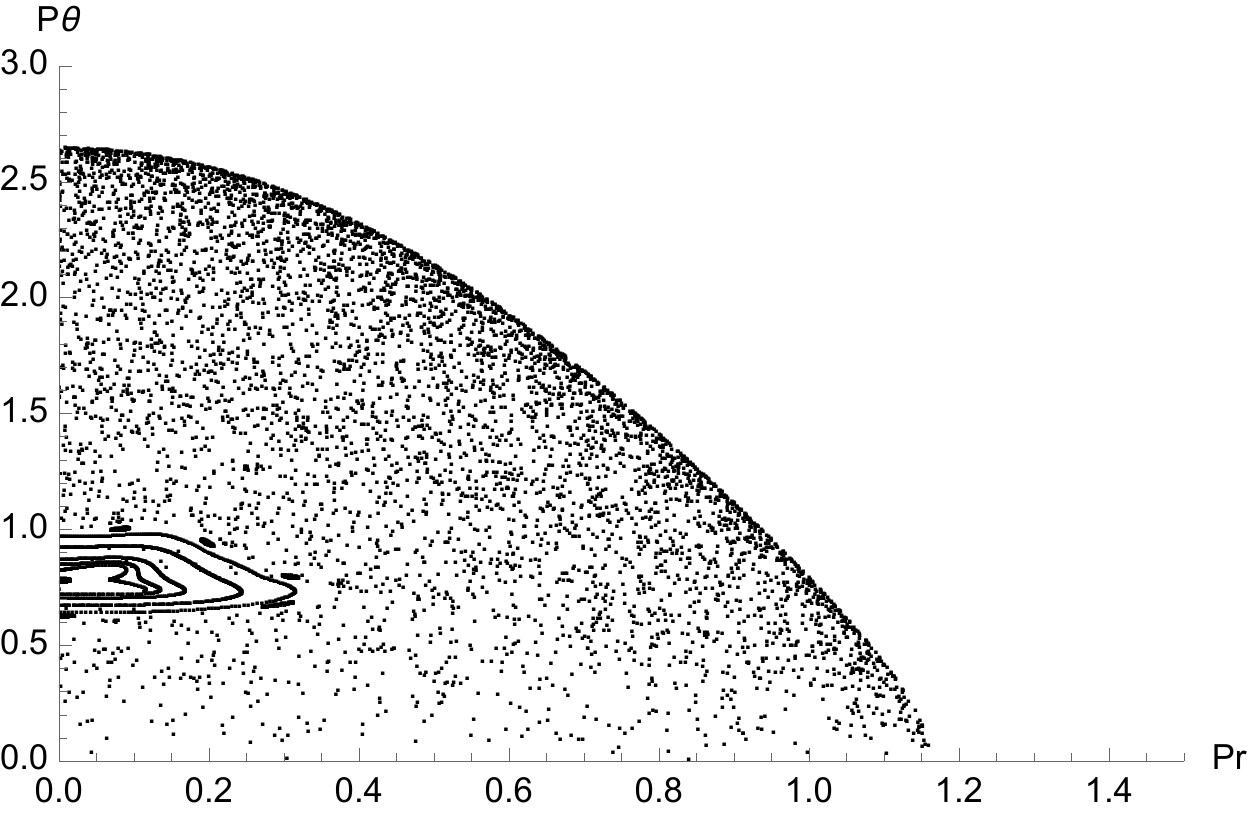}\includegraphics[width=7cm]{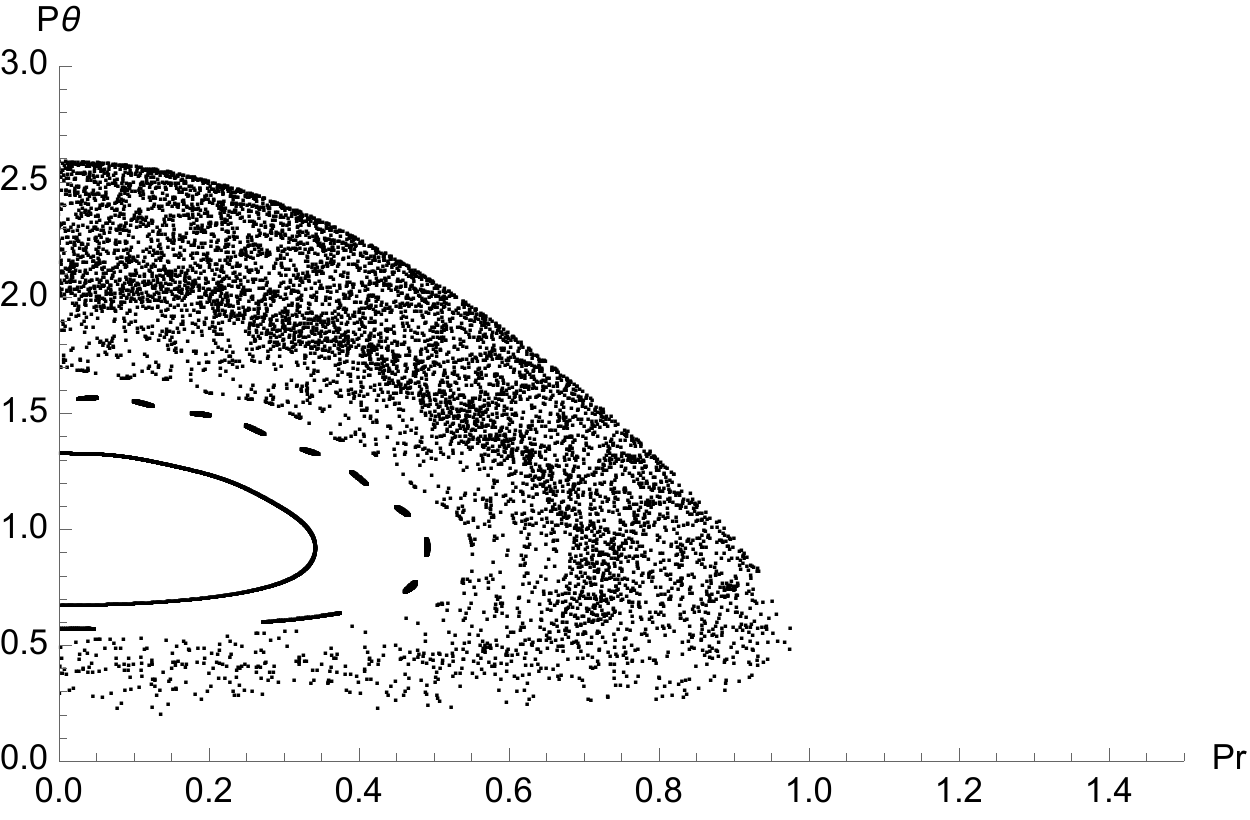}
\includegraphics[width=7cm]{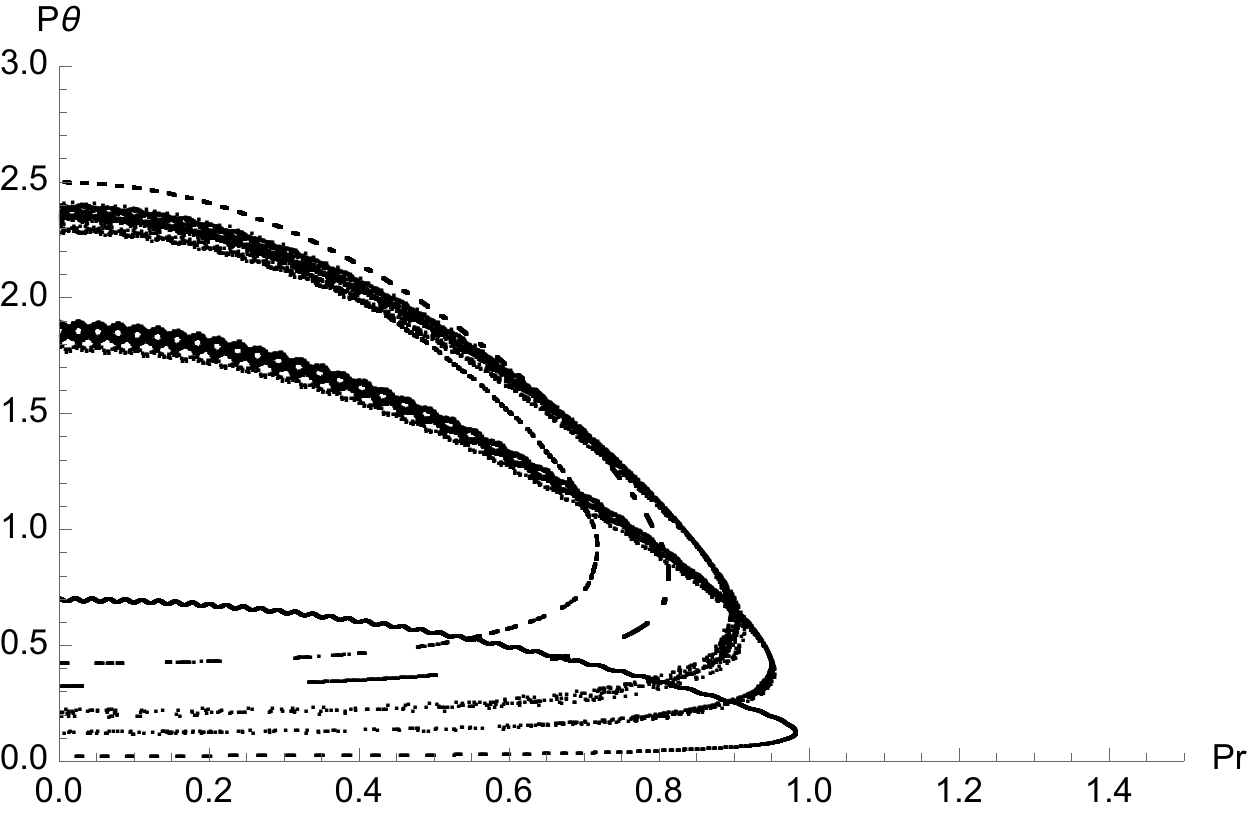}\includegraphics[width=7cm]{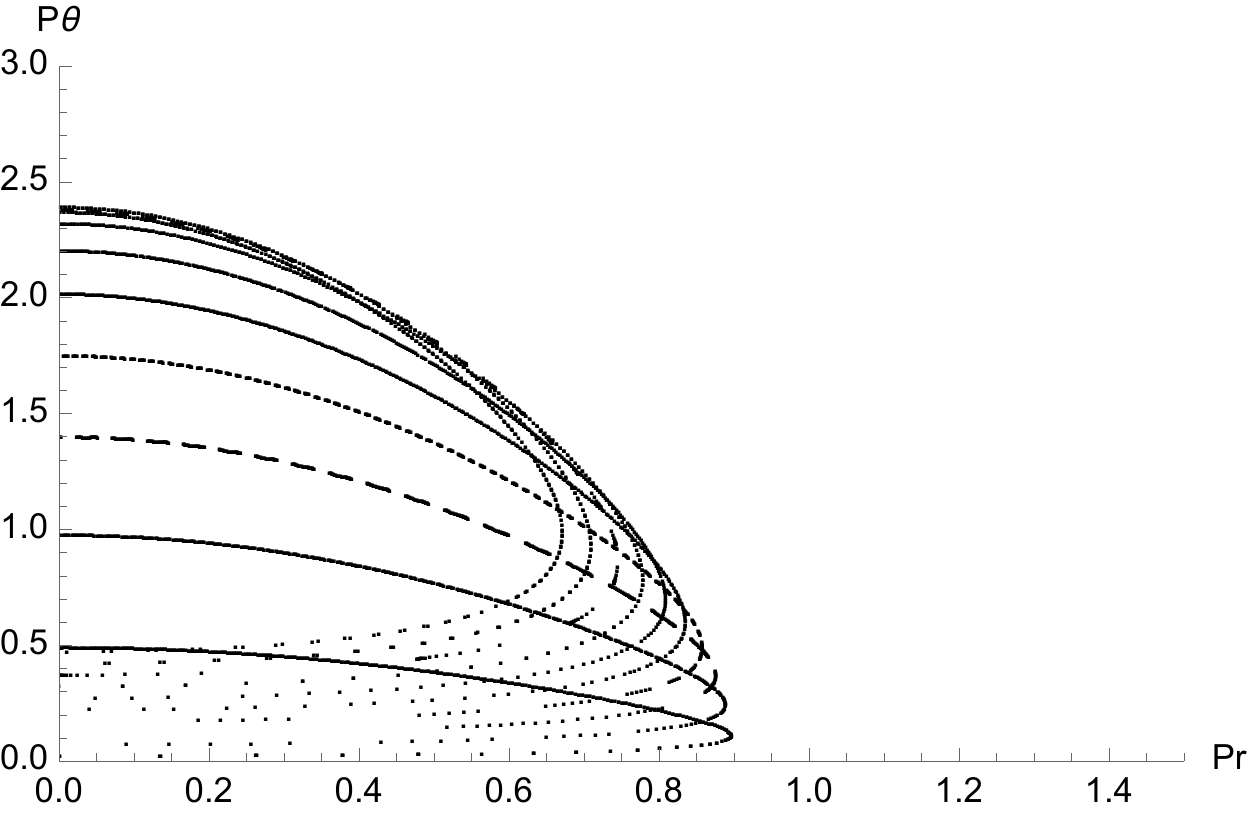}
\caption{Poincar\'e sections at $p_\phi=1, 1.5, 2, 2.5$} \label{fig:islands_stab}
\end{center}
\end{figure}

This is easily visible in terms of a Poincar\'e section of the solution of the dynamics by numerical methods. We choose to take the Poincar\'e section at the crossings of $\theta=0$ in the 
$p_r,p_\theta$ plane. This is shown in figure \ref{fig:islands_stab}.

\begin{figure}[ht]
\begin{center}
\includegraphics[width=7cm]{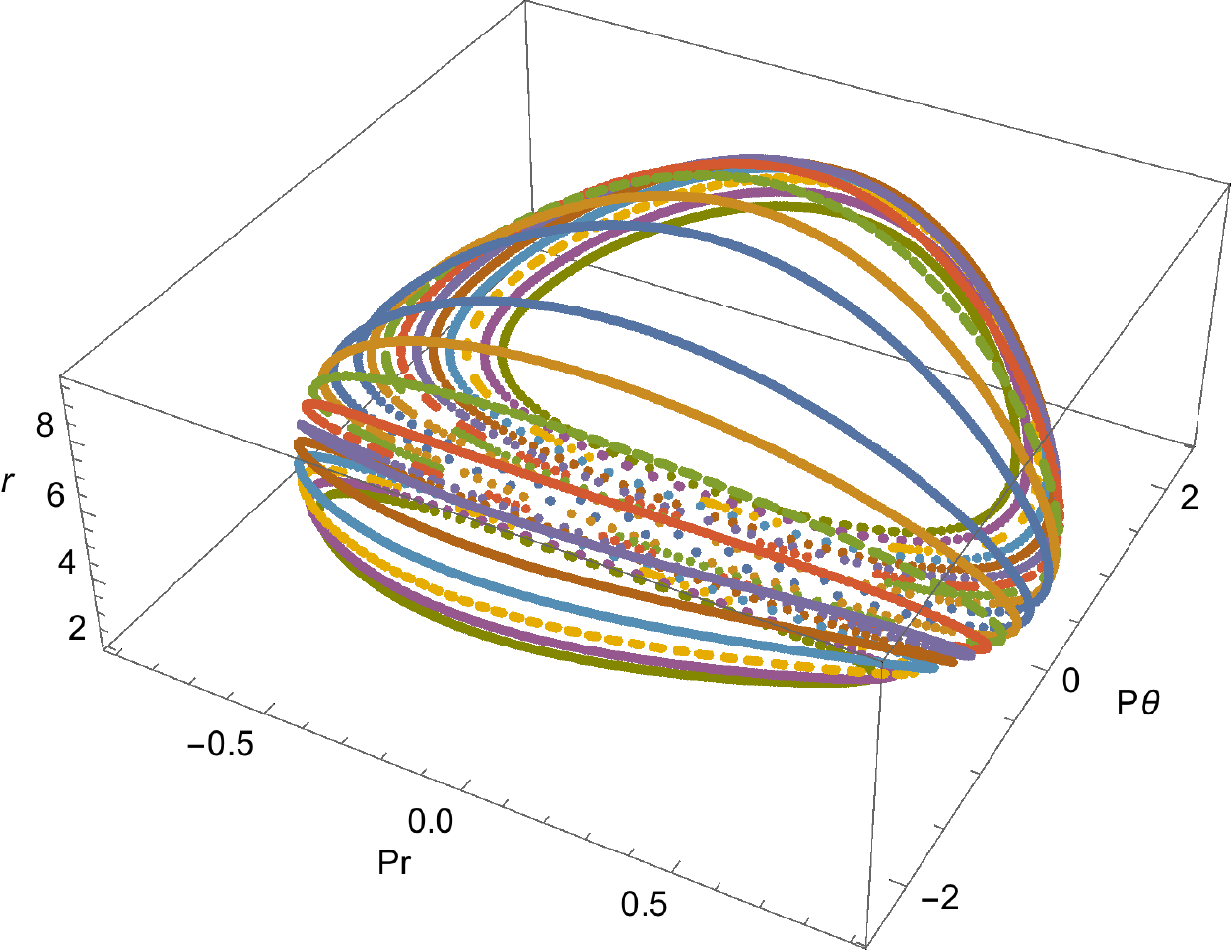}
\caption{Poincar\'e sections at $p_\phi= 2.5$ showing all three $p_r,p_\theta,r$} \label{fig:3dsec}
\end{center}
\end{figure}

As can be seen, KAM tori start forming as we increase the angular momentum $p_\phi$. Their area grows with $p_\phi$ until it seems to take over the available phase space. 
Notice however that the tori seem to intersect, an effect that is especially noticeable in the bottom right corner. This is an artifact of the projection to $p_r,p_\theta$ which ignores the fact that $r$ can have more than one solution when we fix $p_r,p_\theta$ and the energy $E$ at $\theta=0$. This can be seen clearly in figure \ref{fig:3dsec} where we see that different initial conditions (marked with different colors) each gives rise to a pair of circles.

The presence of these topological circles  indicates that in principle there is an additional conserved quantity. That quantity should be thought of as the adiabatic invariant for motion in $\theta$, at least for the small amplitude regime in $\theta$.

Again, as is usual in transitions from integrability (large $p_\phi$) to chaos (small $p_\phi$), the transition happens by destroying some of the KAM tori and then increasing the area of the chaotic region. The important issue for us is that in the classical setup one has to distinguish between different initial conditions in the region of parameter space where there is coexistence between integrable islands and chaos. 

\begin{figure}[ht]
\begin{center}
\includegraphics[width=7cm]{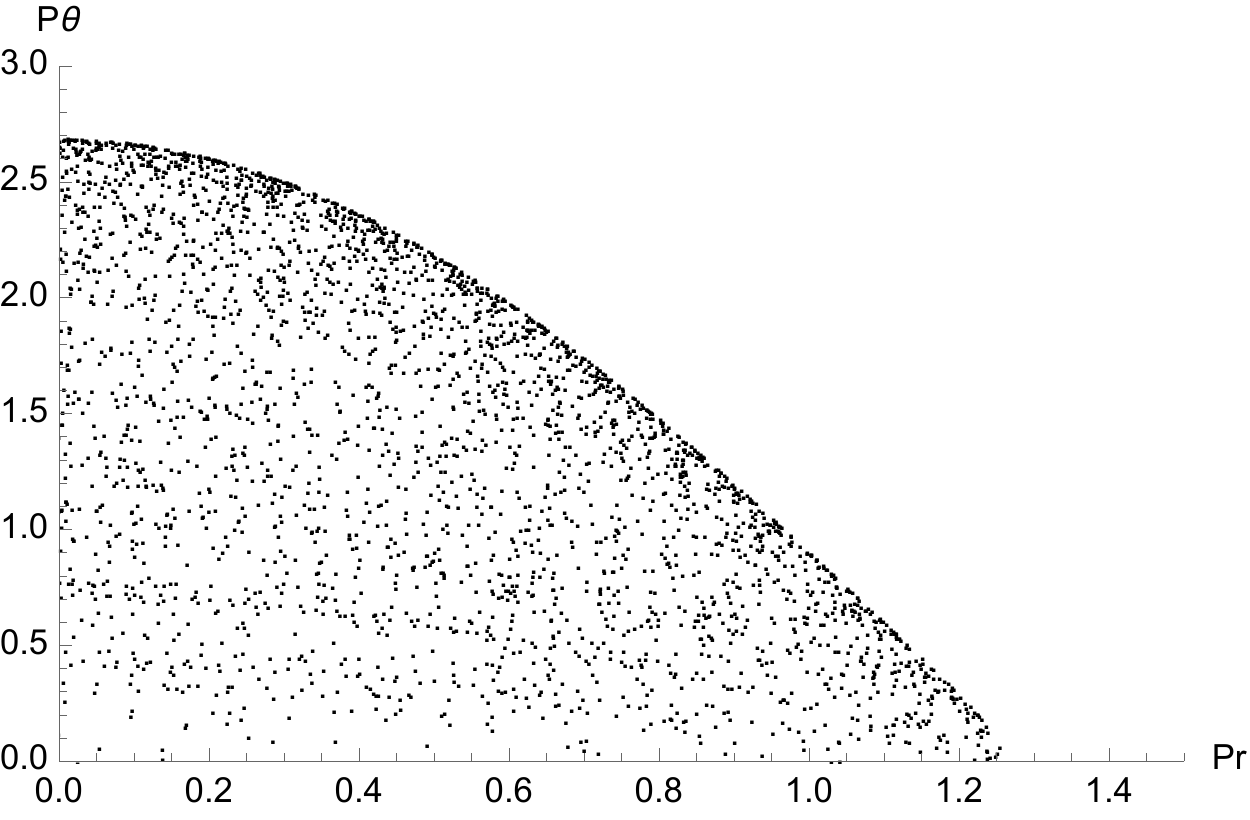}
\caption{Poincar\'e sections at $p_\phi= 0.5$ showing that the chaotic region seems to fill the available phase space.} \label{fig:full_chaos}
\end{center}
\end{figure}

We can check that at $p_\phi=0.5$, the chaotic region seems to have swallowed the whole available phase space. This is shown in figure \ref{fig:full_chaos}

We will label the different regions as phases. Since the matrix model that we are analyzing is closely related to matrix models that describe black holes, we will label the chaotic phase as the black hole  phase. Essentially, such a phase thermalizes over it's available phase space and in larger matrix models has been argued to be related to black holes.  We will label the other phase the orbiting D-brane phase, which can be thought of as a pair of D-branes orbiting each other and having a fixed number of strings stretched between them, with a force that depends on the number of strings that have been excited. This number is the ``adiabatic invariant" for the orbits.
The phases can coexist for some values of $E,p_\phi$ but not others.
From the figures in \ref{fig:islands_stab}, it seems that the coexistence phase disappears somewhere between $2<p_{\phi ,crit}<2.5$ at energy $E=1$. Here we would get a pure orbiting D-brane phase. The coexistence also appears somewhere between $p_\phi=0.5$ and $p_\phi=1$. Below the corresponding value of $p_\phi$, we would call this a pure black hole phase.

\section{Lyapunov exponents and the black hole to D-brane transition}

Now that we have established that chaos can appear and disappear at a particular value of the angular momentum (at fixed energy) it makes sense to try to understand this transition in more 
detail. In particular, one might want to understand to what extent the transition changes the scrambling rate of the dynamics. In this case, the scrambling rate will be captured by the largest (and only) positive Lyapunov exponent. 

We are interested in understanding the chaotic region (numerically) near where it disappears. The edge of chaos in our dynamics is interpreted as the end of the black hole phase. One can think of this limit as an 
extremal limit for a family of black holes. Such extremal limits usually have zero temperature and they lack a horizon, although they still have a near horizon geometry. Recent studies suggest that for black holes, the largest Lyapunov exponent is controlled by the temperature of the black hole system \cite{Shenker:2013pqa,Maldacena:2015waa}. Near an extremal limit, the effective temperature should go to zero and it is reasonable to assume that the corresponding Lyapunov exponent goes to zero near  such a transition. We will present evidence of that effect and we will compute the approach to criticality. Our findings are consistent with a critical exponent of $1$ for the Lyapunov exponent. 

Our simulations are done at fixed energy (set to $E=1$), and a small value of the Planck constant correction ( set to $\hbar=0.1$). Our results are displayed in figure \ref{fig:Lyap} (the table of values from which the plot is extracted can be found in the appendix \ref{app}, table \ref{tab:values}).

\begin{figure}[ht]
\begin{center}
\includegraphics[width=10cm]{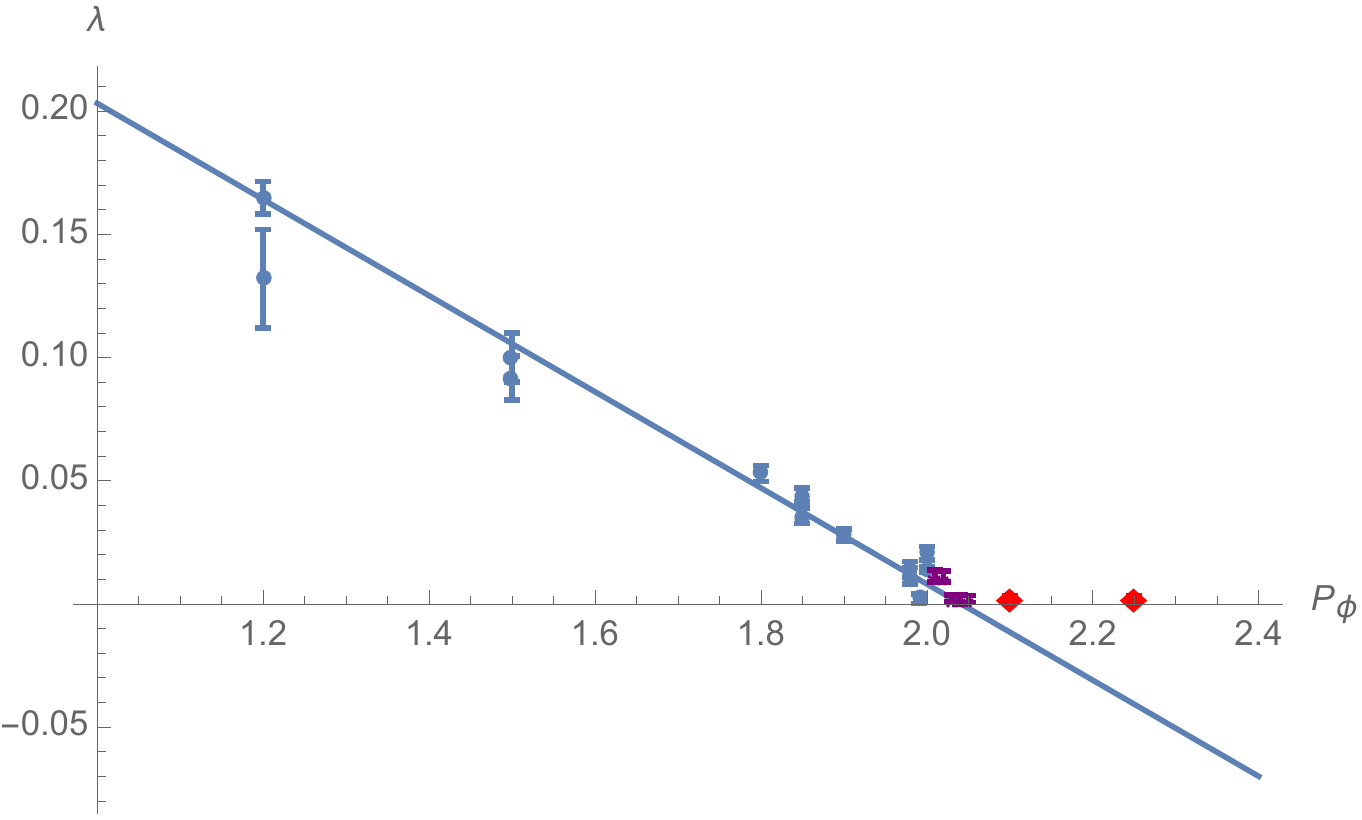}
\caption{Lyapunov exponents for various values of angular momentum, and a linear fit. The bars indicate the statistical uncertainty. Two sets of data were obtained from running the same code with the same data on two  different computers and are superimposed. They are statistically consistent with each other. 
} \label{fig:Lyap}
\end{center}
\end{figure}

\subsubsection{Initial condition and systematics}

The initial conditions for the data are chosen with the following protocol. One first considers using $p_\theta=0, p_r=0$ and $E=1$ in the Hamiltonian \eqref{eq:Hamredcorr}. One varies $p_\phi$ and for each $p_\phi$ one selects a small value of $\theta$. 
 One then computes $r$ numerically. We pick the lowest value of $r$ that is positive. When we trie the other values of $r$ we found that they were usually in the integrable portion of the dynamical system.
 This would give us a starting initial condition that can be characterized as a point on an  equipotential of figure \ref{fig:pot} with zero velocity. The system is then left to evolve for a 
 total run time of $t=5000$. The Lyapunov exponent is computed by following an infinitesimal fluctuation $\delta v$ to the initial conditions in the linearized approximation. 
 For a general Hamiltonian system we do this as follows
 \begin{equation}
 \dot q_i +\delta \dot q_i = \partial_{p_i} H (q+\delta q, p+\delta p)=\partial_{p_i} H (q)+ \partial_{p_i} \partial_{q_j} H(q,p) \delta q_j+\partial_{p_i} \partial_{p_j} H(q,p)\delta p_j
 \end{equation}
 so that 
  \begin{equation}
\delta \dot q_i = \partial_{p_i} \partial_{q_j} H(q,p) \delta q_j+\partial_{p_i} \partial_{p_j} H(q,p)\delta p_j
 \end{equation}
and similarly for  $\delta p$. This is a linear equation for the fluctuation, so we can take the $\delta q, \delta p$ of order one in the numerical simulation.

We evolve the system and record the stretching of the fluctuation at intervals  $\Delta t$, by computing $\lambda_m= \log( |\delta q(m \Delta t)|/|\delta q((m-1) \Delta t)|)/\Delta t$. 
What choice of norm we use matters little (see \cite{Eichhorn} and the appendix A in \cite{Gur-Ari:2015rcq})
We then rescale the $\delta q$ at each such time to start with unit norm. This process is done typically at $\Delta t=500$ and in some cases we take $\Delta t=250$ near the transition in points marked with $x$. The data shown in figure \ref{fig:Lyap} gives the average 
and the variance of different values  (divided by $\sqrt k$, the number of time intervals), which gives the statistical significance of the average result. 

\begin{figure}[ht]
\begin{center}
\includegraphics[width=6cm]{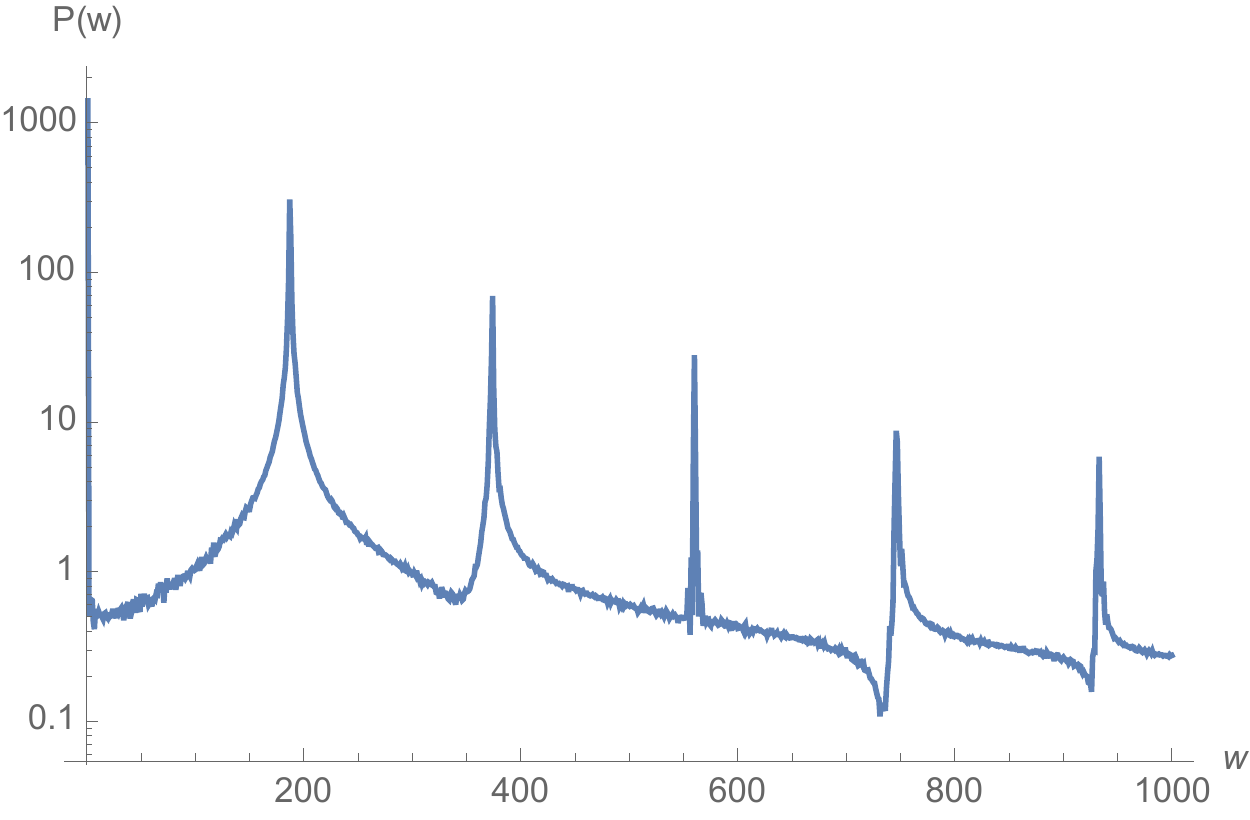}\ \includegraphics[width=6cm]{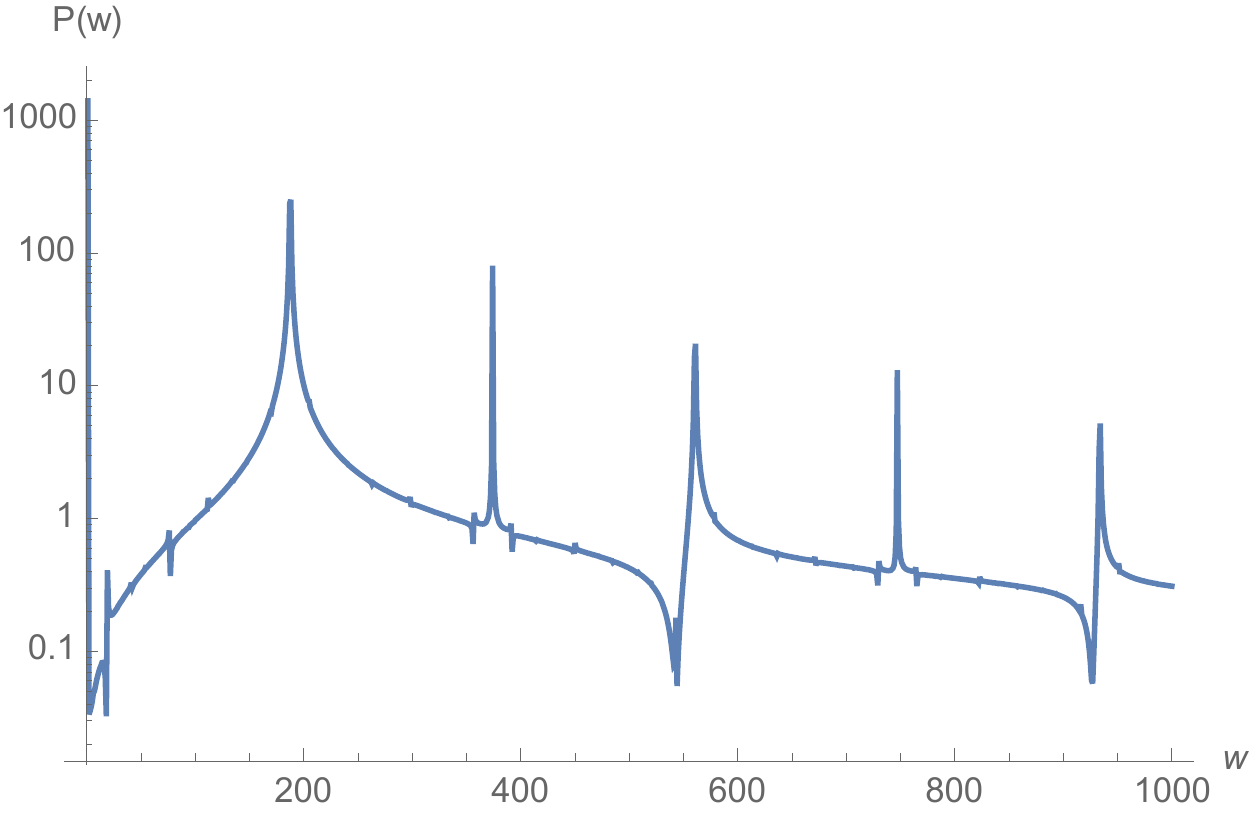}
\caption{Power spectrum for trajectories at $p_\phi=2.02$ and $p_\phi=2.03$ on a logarithmic scale. On the left is the chaotic power spectrum, and on the right is the integrable spectrum. The main difference is the smoothness of the curves and the small addition of small very sharp peaks on the integrable side. 
} \label{fig:intvscha}
\end{center}
\end{figure}

We still need to worry about systematics. In the plot \ref{fig:Lyap} there are points marked with $x$ that are discarded from the fit, but they seem to be consistent with it. Some of these points seem to be in the integrable region. We test this also by studying the power spectrum of the Fourier transform of the solution. Integrable regions have a power spectrum that consists of delta functions 
at the values of the frequencies in the action-angle variables. Similarly, chaotic regions display a continuous power spectrum (such techniques were used to analyze the large $N$ limit of matrix model dynamics in \cite{Asplund:2012tg}). This is shown in figure \ref{fig:intvscha}.
 The ones that are marked with diamonds are consistent with zero and are safely beyond the transition, so they should not be included as part of the fit to chaos close to the transition.

It should be noted that once the Lyapunov exponents are very close to zero, the chaotic region is also shrinking in size, so it becomes harder to hit it with our initial condition choice.  

\section{Conclusion}

In this paper, we studied the transition from chaotic behavior to integrable one in the 2 $\times$ 2 traceless hermitial matrices.
Thanks to the $SU(2)$ gauge symmetry and $SO(2)$ global symmetry, this model can be reduced to two dynamical variables and their conjugate momenta.
The plot of the Lyapunov exponent clearly shows that the intensity of chaos drops as a control parameter increases and goes to zero on the critical point of the transition. 
This coincidence of the point Lyapunov spectrum reaching zero and the critical point of the transition from chaos to integrability is confirmed by power spectrum. 

This is consistent with the claim in the works\cite{Shenker:2013pqa,Maldacena:2015waa} about the end of an extremal black hole.
Namely, the (toy) model we studied shares a black-hole like property with more complex matrix models in spite of the minimality of number of effective degrees of freedom : only 2 dynamical variables and their conjugate momenta.
This simplicity makes the model a desirable laboratory for the study of the relation between matrix models and  gravity.

The next step toward the understanding of the relation between matrix models and gravity will be the study of chaos including quantum effects. It is particularly interesting that because the system is low dimensional, one can directly access the wave functions of the system \cite{Hubener:2014pfa}.
Obviously, it is interestng to understand the relationship between the classical phase diagram and the properties of the quantum wave-functions of the system. It is also interesting to try to compare the classical Lyapunov exponents and their corresponding quantum version to understand better the bounds \cite{Maldacena:2015waa} in an extreme setup that can still be argued to be a holographic model (see also \cite{Berenstein:2015yxu} for arguments that such types of bounds should be generic).

\acknowledgments
D.B. Work supported in part by the U.S. Department of Energy under grant DE-SC0011702. D.B. is very grateful to the Galileo Galilei Institute for their support where part of of this work took place. D.K. is supported by the Japan Society for the Promotion of Science(JSPS). D.K. is very grateful to the grant for the development of international cooperations from the Department of Physics in Kyoto University. D.K. is supported in part by the JSPS Japan-Hungary Research Cooperative Program and the JSPS Japan-Russia Research Cooperative Program.

\appendix

\section{Tables of values}\label{app}

Here we present the data of the two data sets joined. Values where $P_\phi$ appears twice indicate that those two values were run with the same code on different computers.
\begin{table}
\begin{tabular}{|c|c|c|}
\hline $P_\theta$ (Angular momentum) & $\lambda$ (Lyapunov exponent) & Statistical uncertainty\\
\hline 1.2 & 0.132007 & 0.0203839 \\
 1.2 & 0.164935 & 0.0066 \\
 1.5 & 0.091657 & 0.00895991 \\
 1.5 & 0.1 & 0.0099 \\
 1.8 & 0.052924 & 0.00332 \\
 1.85 & 0.0357303 & 0.00318009 \\
 1.85 & 0.0442096 & 0.00276756 \\
 1.9 & 0.02801 & 0.002478 \\
 1.98 & 0.0114 & 0.00336 \\
 1.98 & 0.0139 & 0.00322 \\
 1.99 & 0.00211186 & 0.00195237 \\
 2. & 0.0139116 & 0.000959321 \\
 2. & 0.0206 & 0.00275 \\
  2.01 & 0.0113189 & 0.00244926 \\
 2.02 & 0.0111035 & 0.00215886 \\
 2.03 & 0.0021285 & 0.00146813 \\
 2.04 & 0.00194419 & 0.0018921 \\
 2.05 & 0.00197842 & 0.00148593 \\
 2.1 & 0.001978 & 0.00148 \\
 2.25 & 0.00197842 & 0.00148593 \\
 \hline\end{tabular}
 \caption{Table of values of Lyapunov exponents used for figure \ref{fig:Lyap}}\label{tab:values}
 \end{table}
\pagebreak

\end{document}